\documentclass[12pt,preprint2]{aastex}
\usepackage{graphicx}
\usepackage{fancyheadings}
\usepackage{color}

\newcommand{\gae}{\mathrel{\raise .4ex\hbox{\rlap{$>$}\lower 1.2ex\hbox{$\sim$}}
}}
\newcommand{\pks}{PKS~1127-145}

\shorttitle{Chandra discovery of an X-ray jet in \pks}
\shortauthors{Siemiginowska  et al.}

\begin{document}

\title{CHANDRA Discovery of a 300~kpc  X-ray Jet \\
in the GPS Quasar \pks}

\author{Aneta Siemiginowska$^1$,  Jill Bechtold$^2$ \\
 Thomas L. Aldcroft$^1$,  Martin Elvis$^1$, D.E. Harris$^1$, Adam Dobrzycki$^1$}
\affil{$^1$ Harvard-Smithsonian Center for Astrophysics \\ 
$^2$ Steward Observatory, University of Arizona}

\email{asiemiginowska@cfa.harvard.edu}

\begin{abstract}

\medskip
We have discovered an X-ray jet with Chandra imaging of the $z$=1.187
radio-loud quasar \pks.  In this paper we present the Chandra X-ray
data, follow-up VLA observations, and optical imaging using the HST
WFPC2. The X-ray jet contains 273$\pm 5$ net counts in 27\,ksec and
extends $\sim 30\arcsec$\, from the quasar core, corresponding to a
minimum projected linear size of $\sim 330h_{50}^{-1}$~kpc. The
evaluation of the X-ray emission processes is complicated by the
observed offsets between X-ray and radio brightness peaks.  We discuss
the problems posed by these observations to jet models.  In addition,
PKS 1127-145 is a Giga-Hertz Peaked Spectrum radio source, a member of
the class of radio sources suspected to be young or ``frustrated''
versions of FRI radio galaxies.  However the discovery of an X-ray and
radio jet extending well outside the host galaxy of PKS 1127-145
suggests that activity in this and other GPS sources may be long-lived
and complex.

\smallskip

\end{abstract}

\section{Introduction}

The unprecedented sub-arcsecond resolution (Van Speybroeck et al,
1997) of the {\it Chandra X-ray Observatory} (Weisskopf \& O'Dell
1997) gives us, for the first time, the opportunity to study details
of the X-ray structures associated with distant galaxies and quasars.
Jets observed in Galactic and extragalactic sources are still not well
understood and before the {\it Chandra} launch only a handful of
nearby sources were known to have X-ray emission associated with their
radio jets (e.g. Harris 2001).  {\it Chandra} has uncovered many X-ray
jets with complex structure on arcsec scales (e.g. Marshall et al
2001a,b, Chartas et al 2000, Schwartz et al 2000) in sources up to
$z$=0.6.

\pks \, is a higher redshift quasar ($z$=1.187) with a GigaHertz
peaked radio spectrum (GPS, Stanghellini et al., 1998), intervening
damped Lyman-$\alpha$ (Bergeron \& Boisse, 1991) and HI 21-cm
absorption, both at redshift z=0.312 (Lane et al., 1998).  We observed
\pks\ with the {\it Chandra} X-ray Observatory, in order to study in
detail the quasar X-ray spectrum and the nature of the absorber
(Bechtold et al. 2001), but were immediately struck by the presence of
a 30$\arcsec$ long, one-sided X-ray jet.  At the quasar redshift a
30$\arcsec$ separation implies that the X-ray emitting jet material is
at least $\sim 250-330h_{50}^{-1}$~kpc from the central engine which
creates the jet.  If the jet is not in the plane of the sky the
physical distance will be even larger, quite possibly a Megaparsec or
more.

The radio spectrum of the core of PKS 1127-145 places it in the GPS
class of radio sources (O'Dea 1998) and the extended radio structure
in \pks\ is weak, only $\sim$0.1-0.4$\%$ of the core emission
(Sec.3.2).  This is much less than a typical large scale emission of
FRI or FRII galaxies, where the lobes can dominate or be comparable to
the core brightness (Kellerman \& Owen 1988).  Because GPS sources
generally show very compact radio morphology ($<$~1kpc), they have
been interpreted as either young counterparts of FRI radio galaxies
(Phillips \& Mutel 1982), or as ``frustrated'' AGN (van Breugel 1984,
O'Dea 1998), in which the radio jets are not able to penetrate the
host galaxy's gas and dust. There are just a few examples of the GPS
galaxies with a very faint Mpc scale radio structures (Schoenmakers
1999), which are interpreted as relics of the source past
activity. \pks\ is the first example of a GPS quasar with large scale
X-ray emission associated with the faint radio structure.

In this paper we present the {\it Chandra} X-ray imaging data, as well
as the results of our follow-up VLA and HST/WFPC2 observations of
\pks. We then discuss the jet morphology and possible emission
processes, and conclude by considering implications of the large scale
X-ray jet on the origin and evolution of GPS sources.

The main results of our observations are (1) the discovery of the
large scale X-ray jet in a GPS quasar; (2) improved morphology and
frequency coverage of the weak radio jet; and (3) the detection of
displacements of the peak brightnesses between the radio and X-ray
emission in the knots, such that the X-ray precedes the radio emission
moving outward along the jet.

We assume H$_0$=50~km~s$^{-1}$~Mpc$^{-1}$, q$_0$=0 throughout the
paper, so 1$\arcsec$ corresponds to 11.5~kpc at the quasar redshift.

\begin{figure}
\epsscale{0.85}
\plotone{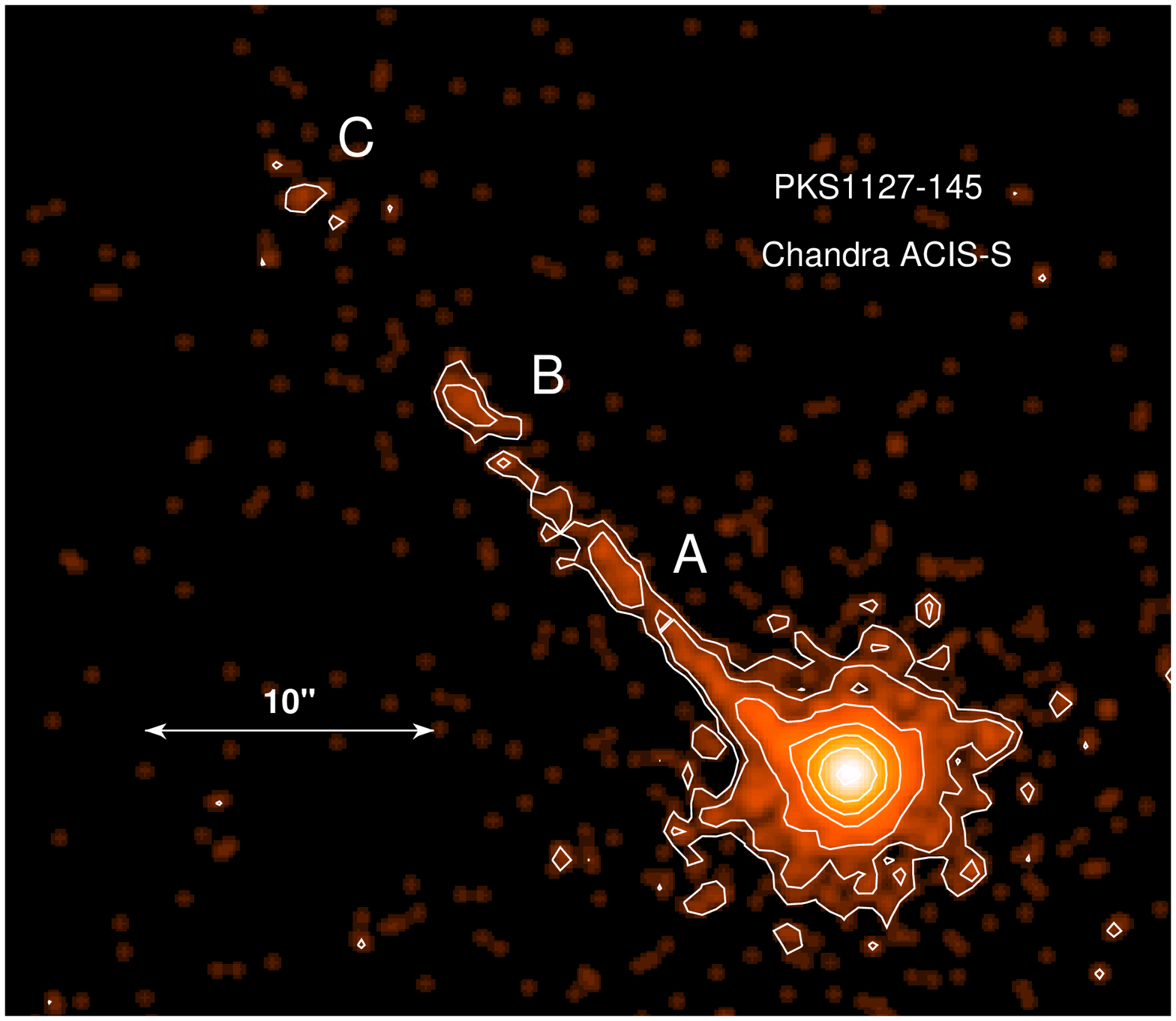}
\caption{\small PKS~1127-145 ACIS-S image 
smoothed with a Gaussian kernel ($\sigma = 0.25\arcsec$).  {\it
Chandra} background is increasing rapidly at high energies, therefore
only the events with energies between 0.3-6~keV were included in this
image.  North is up and East is left. The knots are labeled
A,B,C. Contour levels: 1.2,2.6,4.5,9,27,3000 counts/pixel.}
\end{figure}

\section{The Observations}

\subsection{{\it Chandra} X-ray Data}

We observed \pks\ for 27,358 seconds with the spectroscopic array of
the Advanced CCD Imaging Spectrometer (ACIS-S, Garmire et al, in
preparation; Weisskopf et al. 1996) on 2000 May 28 (ObsID 866) without
any transmission grating in place.  The source was located on the back
illuminated chip (S3) and offset by $\sim 35\arcsec$ from the default
aim point position to avoid the node boundary ({\it Chandra} Proposers'
Observatory Guide, POG, 2000).  The 1/8 subarray mode with a 0.43
second frame time was chosen to mitigate pile-up, leading to a
1$\arcmin$x8$\arcmin$ image.  The X-ray position of the quasar (J2000:
11:30:07.03, -14:49:27.32) agrees with the optical position (Johnston
et al 1995) to better than 1$\arcsec$, as expected given the quality
of the {\it Chandra} aspect solution (Aldcroft et al 2000).  We used
CIAO (version 2.1; http://asc.harvard.edu/ciao/) software to analyze
the data.

A smoothed image of \pks\ and its associated jet is shown in Figure~1.
The intensity has been scaled logarithmically to emphasize the faint
jet emission.  The small {\it Chandra} Point Spread Function (PSF),
especially the low power in the scattering wings ($r>1\arcsec$),
allows for a high dynamic range image, which is essential to the
detection of the jet (Figure 1).  A total of 16,573 counts was
detected in the quasar core, so the total jet emission is
$\sim 60$ times fainter than the core (see Section 3.1.3) and the
individual components are as weak as $\sim 1/450$ times the core. The
details of the spectral analysis of the central source are reported
elsewhere (Bechtold et al. 2001).  {\it Chandra} ACIS-S3 data were
reprocessed with the pipeline version R4CU5UPD13.3 on Jan.24, 2001.
We have used the reprocessed data and calibration files available in
CALDB v.2.1. We note that the aspect uncertainty on the absolute
position is less than $\sim 0.5\arcsec$ and the uncertainty on the
image reconstruction is less than $\sim 0.1\arcsec$ (Aldcroft et al
2001). The PSF FWHM at the quasar core is about $0.75\arcsec$.

In addition to the standard CXC processing we have corrected the
events in the ACIS readout streak ({\it Chandra} POG, 2000) using the
CIAO {\tt acisreadcor} tool assuming a uniform background over the
entire chip. This correction moves the events in the readout streak
into the core region and does not affect the jet analysis. The
corrected image was used only to create a smoothed version of the
image. We used the original event file for the analysis of the jet
emission and morphology.

We have used the readout streak photons to estimate the pile-up
fraction in the data. We have extracted readout streak photons
assuming two box regions 4 pixels wide along the streak, excluding the
circular region with 6.5$\arcsec$ radius centered on the core.  The
count rate in the readout streak 
is 0.073 cts/frame which gives a pile-up fraction of 2-3$\%$ in this
observation ({\it Chandra} POG, 2000).

\subsection{VLA Radio Data}

At the time of the Chandra discovery of the X-ray jet in \pks, the only
reported radio detection of the weak kpc jet was that in the thesis
of Rusk (1988).  Archival VLA data consisted mainly of short
observations (PKS1127-145 is a standard VLA calibrator source) and
maps from these data did not have sufficient dynamic range to detect
the kiloparsec jet.  VLBI results for the core have been published by Wehrle
et al. (1992) and further monitoring data have been accumulated at 2~cm
by Kellermann et al. (http://www.cv.nrao.edu/2cmsurvey/).  The
Brandeis group has also recently obtained VLBA polarization maps at
8.4 GHz  (Homan and Wardle, private communication).  All
these data show a quasi equal E-W double separated by about 4~mas plus
weaker structure in a jet bending off to the NW with a total extent of
about 20 mas.

Our new data were obtained at 1.4 and 8.4 GHz at the VLA\footnote {The
National Radio Astronomy Observatory is a facility of the National
Science Foundation operated under cooperative agreement by Associated
Universities, Inc.} on 2001~Feb~6-7 (program AH730).  The observations
were made in the BnA configuration for 8 hours with on source times of
2.1~hr at 1.4 GHz and 3.7~hr at 8.4 GHz, and 50~MHz bandwidth.  The
source is strong and residual baseline errors may degrade performance.
Two scans of 3C286 were obtained for the flux density scale
(S(1.4)=5.46 Jy and S(8.4)=3.426 Jy) and polarization position angle
calibration and 3C138 was observed once to check the polarization
solution: position angles were within 6$^\circ$ of the expected
value. The rms errors in our reported measurements are equal to
0.12~mJy, which is higher than the theoretical errors for each band.

Although we observed a secondary phase calibrator for backup, our
reductions relied on self calibration since the unresolved core is of
order 100 times brighter than peak intensities of the kpc jet.  The
final pass of {\tt selfcal} included clean components from the knots as well
as the core and reached dynamic ranges (except for the regions
directly north and south of the core) of 20,000 (1.4 GHz) and 65,000
(8.4 GHz).

\begin{figure}
\epsscale{0.85}
\plotone{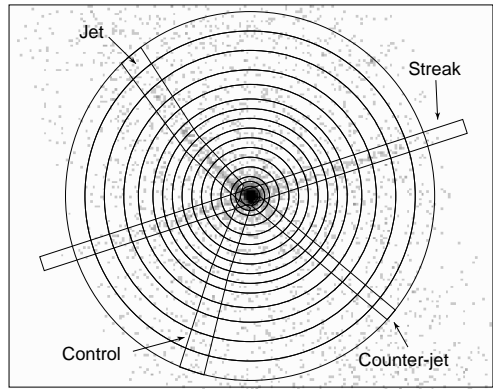}
\caption
{\small ACIS-S X-ray image of PKS~1127-145 (0.3-10~keV) with the
quasar at the center.  
The outermost circle has a radius of 37$\arcsec$. The readout
streaks are visible on both sides of the central core.  The regions
used for calculating the jet profile (polygons along the jet), the
counter-jet (box in the direction opposite to the jet), the background
(annuli excluding the jet and the streak) and the control region
(southern polygons) are overlayed on the ACIS-S events.}
\end{figure}

\subsection{HST/WFPC2 Observations and Data Reduction}

We observed \pks\ and its environs with the {\it HST} WFPC2 for a
total of 8800 seconds.  The first half of the data were obtained on
2001 May 23, and the other half of the data on 2001 August 1.  The
F814W filter, which approximates the Johnson-Cousins I-band filter,
was selected to optimally detect galaxies in the vicinity of the
bright quasar.  \pks\ was placed near the center of the WF3 camera.
The images were calibrated by the HST standard processing pipeline,
including the standard reduction steps of bias subtraction, dark
current subtraction, and flat-fielding (Holtzman et al. 1995).

The eight raw images were combined into a final clean image using standard
tasks within IRAF.  First, the images were aligned to a common roll angle
using the task MSCIMAGE in the MSCRED package.  This is necessary because the
observations were taken at two different epochs with substantially different
roll angles.  Next, the images were combined and cleaned of cosmic rays using
IMCOMBINE.

\section{Analysis}

\subsection{X-ray Jet}

\subsubsection{X-ray jet Morphology}

The jet-like emission at PA$\sim 43^{\circ}$ extends up to $\sim
30\arcsec$ away from the core.  The outer jet does not point directly
at the core, but instead the X-ray jet changes its position angle starting
at $\sim$64$^{\circ}$ 2$\arcsec$ from the core
and ending with PA=43$^{\circ}$ at the outermost knot 27$\arcsec$
from the core.

Figure 2 shows the central (80$\arcsec$x80$\arcsec$) of the event file
with 2 prominent readout streaks in position angles PA$\sim$110/290
degrees.  We extracted radial profiles for the jet and possible
counter-jet using the regions shown in Figure 2. For test purposes we
also extracted a control profile with the same geometry as the jet,
but at a position angle expected to be free of jet emission and
instrumental features.  The profiles are shown in Figure~3.
Continuous jet emission is detected to about 22$\arcsec$ from the
core, where it drops and then reappears at $\sim 25.5 \arcsec$
distance.  Three knots can be distinguished in the jet: A($\sim
10\arcsec$), B($\sim 18\arcsec$) and C($\sim 27\arcsec$ from the
core). The jet is resolved transversely with a size of order
1.5$\arcsec$-2$\arcsec$ and the size of the knots along the jet
2.5$\arcsec$-3.5$\arcsec$.  The knots correspond roughly to the VLA
radio knots, but the peak brightnesses of the radio and X-ray emission
do not match exactly (Sec.3.4).

The physical distance from the nucleus to the end of knot B is
about 255$h_{50}^{-1}$ (sin$\theta)^{-1}$~kpc,
where $\theta$ is the jet angle to the line of sight.  The physical
distance from the nucleus to the outer knot C is 
$\sim 330h_{50}^{-1}$ (sin $\theta)^{-1}$~kpc.

There is no detection of a counter-jet.  Using the observed counts in
the jet, counter-jet and background regions we derive a lower limit of
5:1 (90$\%$ confidence) on the flux ratio between jet and counter-jet.
This indicates that we see the jet at an angle smaller than 11$^\circ$
(assuming normal relativistic beaming, Rybicki \& Lightman 1979). This
corresponds to a jet of $>$1.5~Mpc. This estimate is based on the
assumption that the jet is straight, however, \pks\ jet curves and the
above number should be treated as an order of magnitude estimate. If
the angle is larger at large distances then the jet overall length is
smaller.

It should be noted that the counter-jet profile between
2$\arcsec-6\arcsec$ from the core shows evidence for a non-zero flux,
in contrast to the central region. The likely explanation is a
presence of a weak extended component on this scale (see Sec.3.3).

\begin{figure}
\epsscale{1.0}
\plottwo{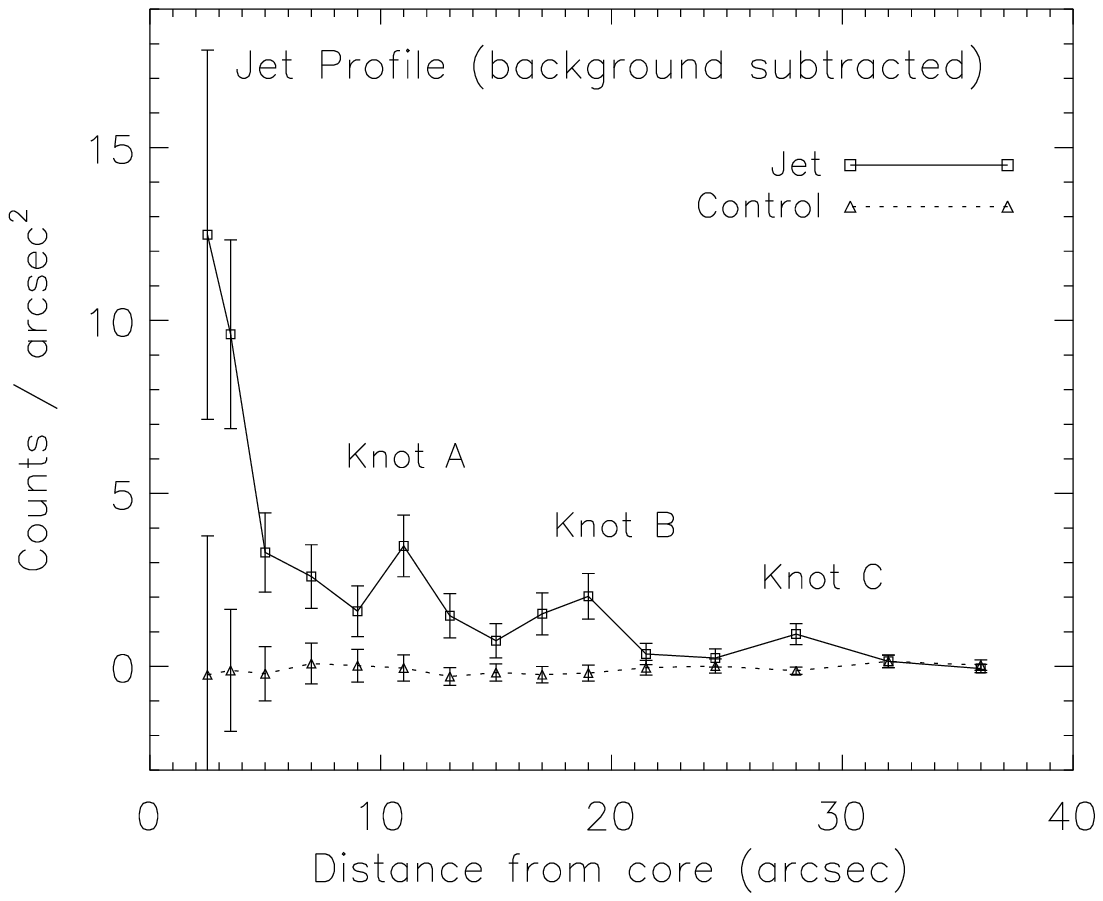}{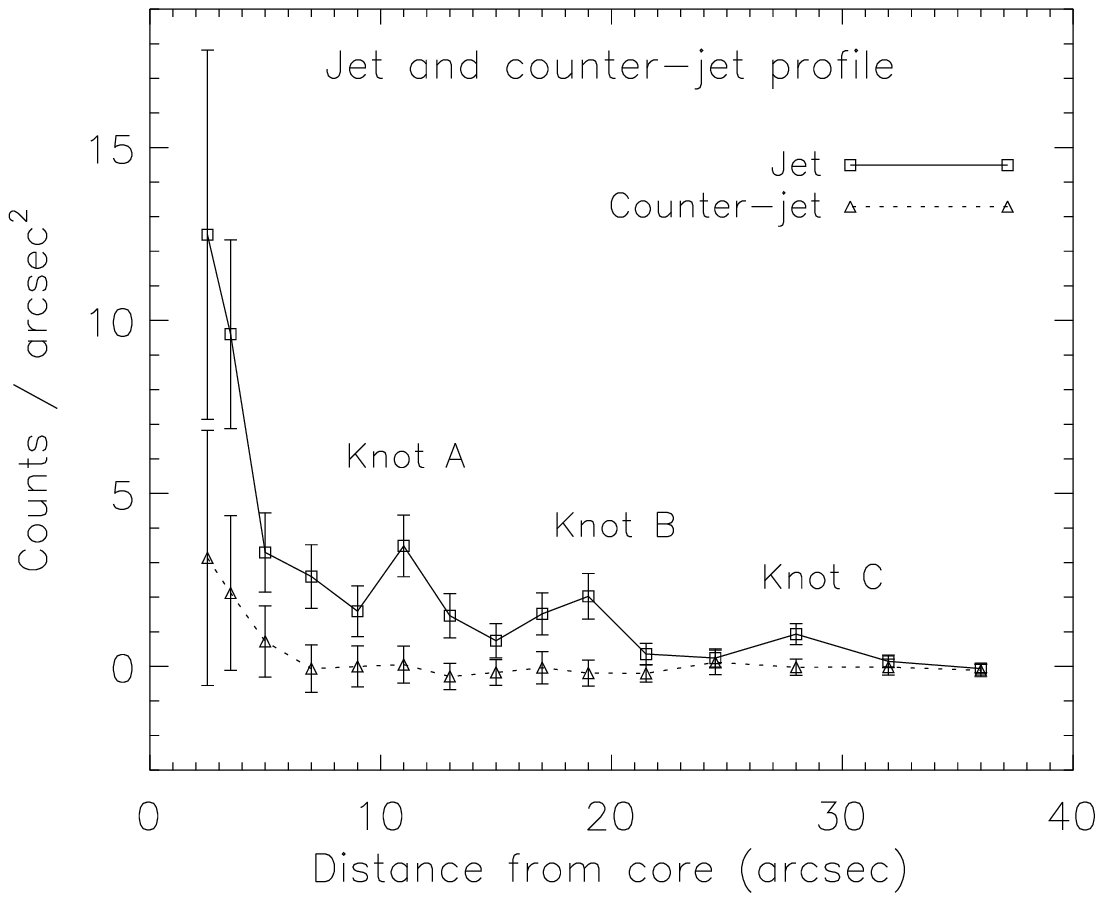}
\caption
{\small Jet profiles. Left panel: The background subtracted jet
profile, calculated using polygon regions along the jet, is indicated
with the squares.  The background was extracted from the annuli
centered on the core, excluding the jet and the streak region
(Fig.2). The control profile extracted from the southern part of the
quasar is plotted with triangles and connected with the dashed line.
Only photons within 0.3-10 keV energy are included in the
profile. Right panel: Jet and counter-jet profiles. No counter-jet is
detected.}
\end{figure}

\subsubsection{X-ray Jet Intensity \& Spectrum}

We have extracted 368 total counts (with {\tt dmextract}) from the
{\it Chandra} event file (evt2) using the polygonal region which
starts 2$\arcsec$ from the central core and contains the jet as shown
in Fig.2.  The background was extracted using the largest annulus
(with inner and outer radii equal to 3$\arcsec$ and 37$\arcsec$
respectively) excluding the core, the readout streaks and the jet
polygon. Instrument response files were calculated using {\tt mkrmf
and mkarf} tools and CALDB 2.1.  We fit the jet spectral data in {\it
Sherpa} (CIAO2.1) including only the energy range within 0.3-7~keV. We
excluded the energies above 7~keV where the ACIS background rises (see
calibration information on the CXC Web page
http://asc.harvard.edu/cal/).  We assumed a simple power law model
with fixed Galactic absorption ($N_H^{gal}=3.89 \times
10^{20}$cm$^{-2}$, Murphy et al 1996) to fit the data. The best fit
photon index was $\Gamma=1.5\pm 0.2$ with a total flux over 2-10~keV
range of $5.2{\pm 0.3} \times 10^{-14}$~erg~cm$^{-2}$~sec$^{-1}$. We
fit the data and the background simultaneously using the C statistic
(Cash 1979), giving a statistic of C=1123.1 for the best fit
model. This corresponds to $\chi^2$ of 62.2(48~d.o.f.) for binned
data.

The corresponding isotropic luminosity within the 2-10~keV band is
equal
to $3.7 \times 10^{44}$ erg~sec$^{-1}$. 

We have also extracted counts for the three individual knots assuming
elliptical regions. The ellipse sizes increase outwards as the knot
emission becomes more extended: knot A:1.5$\arcsec$x2.6$\arcsec$, B:
1.7$\arcsec$x3.3$\arcsec$, C: 2.$\arcsec$x3.8$\arcsec$. The total and
net counts as well as hardness ratios for each knot are presented in
Table 1.  We also calculated the 2-10~keV flux for each knot assuming
a fixed power law index of 1.5 and fit for normalization (Fig.4).  The
flux for each knot decreases with the distance from the core.

\subsection{Radio Emission.}

\subsubsection{Radio Morphology}

The basic structure of the radio emission is shown in Figure 5, which
is a grey scale image of the 8.4 GHz map together with superimposed
contours from the 1.4 GHz data.  At our resolution, we distinguish 3
brightness enhancements (`knots') in the jet-like structure.  As in
the X-ray data, we denote the well separated knots as A, B and the
outer one as C.  At 8.4~GHz we also detect a short section of the
inner jet which can be traced from 1$^{\prime\prime}$ to
2$^{\prime\prime}$ from the core (see Section 3.3).

At 1.4 GHz, all three knots are resolved and typical deconvolved
dimensions from 2D Gaussian fitting in AIPS are of order
3$^{\prime\prime}$ to 5$^{\prime\prime}$ along the jet and
1$^{\prime\prime}$ to 2$^{\prime\prime}$ in the transverse
direction. Knot A is the weakest of the three knots.  Knots B and C
are much stronger and more diffuse.  Overall, it is evident that the
radio jet is curved (PA changes from 64$^\circ$ at the core to
43$^\circ$ at knot C), in a way closely similar to the X-ray jet.

At 8.4~GHz the structures of knots B and C are well resolved and show
complex morphology, but knot A is not detected at this frequency. The
two outer knots are well separated from one another and there is no
detected emission connecting the knots with the core. Because the
physical size of the resolved knots is of order 20-30~kpc they could
be lobes rather than jet knots. Deeper observations at this frequency
or higher resolution data at lower frequency are needed to
discriminate these possibilities.  Note also that usually two lobes
are observed on opposite sides of the core, but we do not see any
radio structures on the other side.  In the following analysis we
assume that the observed radio structures are features of the jet.

\subsubsection{Radio Intensities}

Flux densities were measured by fitting the peak intensities of
features, by integrating in a box around each feature, and by
performing a 2D Gaussian fit.  The values presented in Table 1 are
integrated flux densities primarily based on the box measurements as
the best measure of the actual integrated intensity.  The box sizes
are given in Table 1.

\begin{figure}
\epsscale{0.85}
\plotone{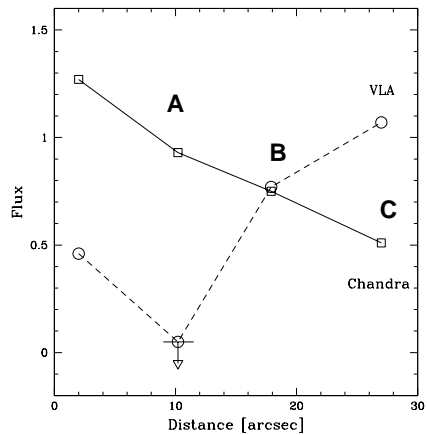}
\caption{\small The X-ray flux and radio flux densities as a function of 
the distance from the core. The X-ray (2-10~keV) flux is plotted in
units of 10$^{-14}$ ergs cm$^{-2}$ s$^{-1}$ and 8.4~GHz flux density
is in units 10~mJy (10$^{-25}$ ergs cm$^{-2}$ s$^{-1}$ Hz$^{-1}$)}
\end{figure}

The values in Table 1 show that the radio intensities of the knots
increase outward, in contrast to the X-ray intensities which decrease
outward (see also Figure 4). This behavior was also observed in the
3C 273 jet (Marshall et al. 2001).

\begin{figure}
\epsscale{0.85}
\plotone{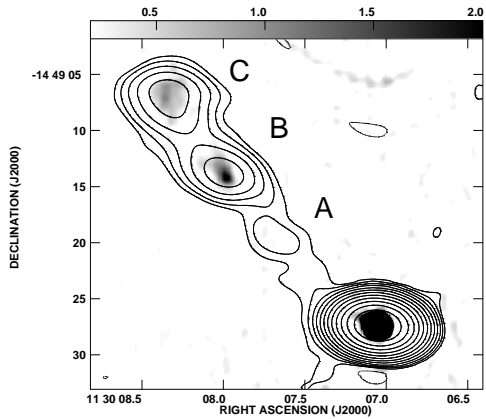}
\caption
{\small 
The VLA BnA array 8.4 GHz data (grey) and 1.4 GHz data (contours).
The grey scale (0.2 to 2 mJy/beam) is a tapered version of the 8.4 GHz
data with a beam of 0.87 x 0.83$^{\prime\prime}$, PA=54$^{\circ}$.
The peak brightness is 3.427 Jy and the rms noise level is 0.12 mJy.
The contours show the 1.4 GHz map with a beam of 3.52 x 2.31
$^{\prime\prime}$, PA=84$^{\circ}$.  We did not use any taper since
the imposition of a circular beam begins to cause confusion amongst
the knots.  The contour levels are logarithmic, increasing by factors
of two and start at $\pm$~1 mJy/beam.  The peak brightness is 5.463 Jy
and the rms noise level is 0.35 mJy.}
\end{figure}

\begin{figure}
\epsscale{0.85}
\plotone{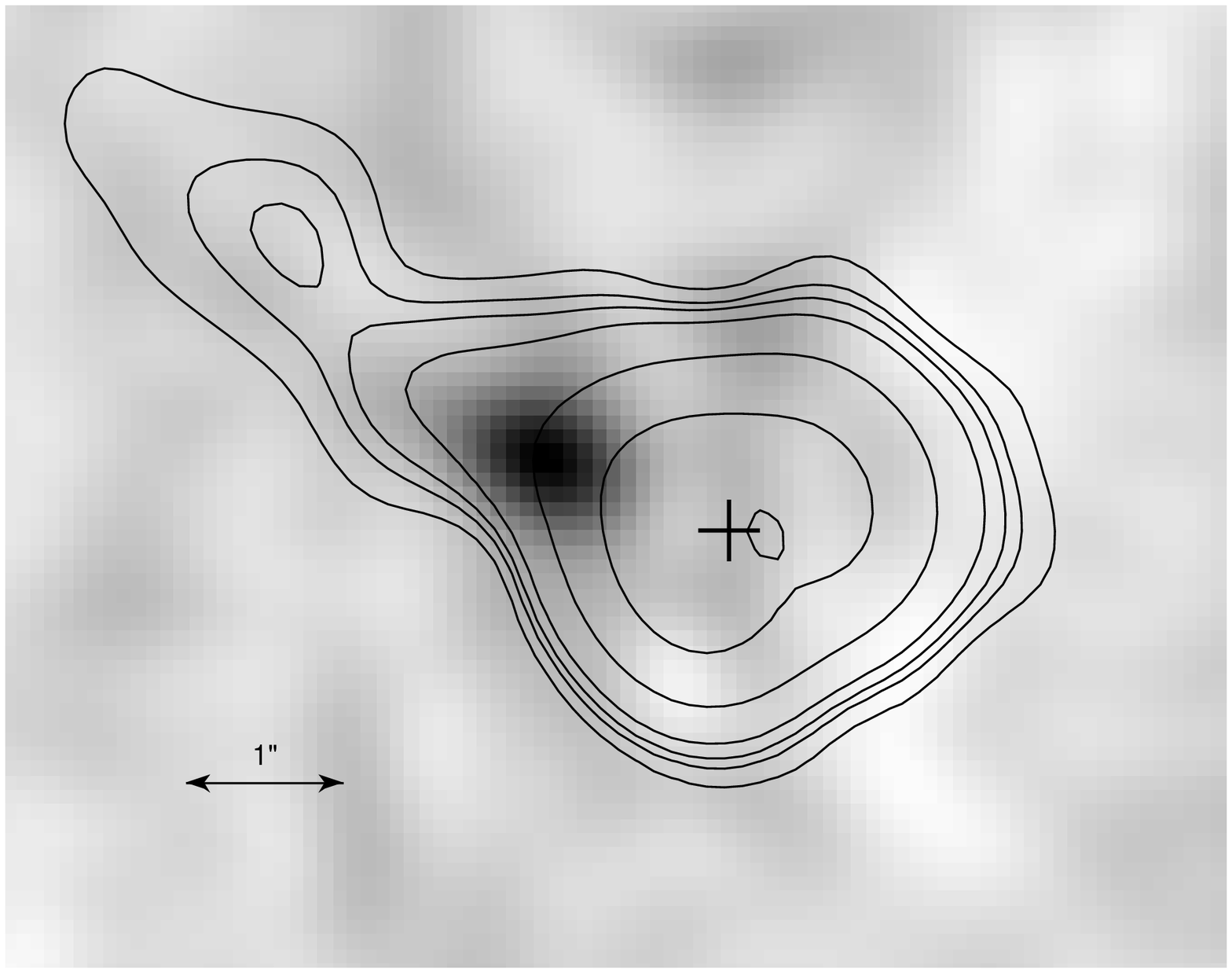}
\caption{\small  Overlayed X-rays (contours) and 8.4GHz (grey
scale) residuals after subtraction of the quasar core emission in the
central region (7.6$\arcsec$x6$\arcsec$). Only X-ray photons in the
energy range between 0.3-6.5~keV were included in the analysis. The
location of the core is marked with a cross. The radio beam size is
0.63$\arcsec$x0.37$\arcsec$ in PA = 77$^{\circ}$}
\end{figure}

\subsection{X-ray structure in the Central 3$\arcsec$ region}

We have analyzed the central $3\arcsec$ region of \pks\ and searched
for emission from the inner jet corresponding to that seen in the
radio. We constructed the PSF at the source location using the PSF
library (F1 in CALDB 2.1).  We also extracted a calibration source
from the archive, the quasar PG~1634+706 (ObsID 69), with similar
X-ray spectral properties, but without obvious extended emission. We
scaled and subtracted the PSF and the quasar profiles from \pks\ core
(using both CIAO and AIPS) and analyzed the residuals for any
significant excess.

Figure 6 shows the contours of the X-rays residuals, after subtracting
the PSF from the core, overlayed on the grey scale of the radio (8.4
GHz) residuals. The excess of the X-ray emission, located $\sim
1.5\arcsec$ from the core at PA=64$^\circ$, corresponds to the excess
of the radio emission.  The PA is consistent with the jet emission
seen on the VLBI scale (Jorstad et al 2001a, 2001b).  The X-ray excess
contains 56 net counts, after subtracting a background extracted from
an annulus with radii between 1$\arcsec$-2.2$\arcsec$ and excluding a
circular region at the inner knot.  The inner knot spectrum is soft
compared to the outer knots (although the uncertainties are larger)
with a best fit photon index $\Gamma = 2.0\pm 0.5$ corresponding
to a 2-10~keV flux of 5.26$\times 10^{-15}$~ergs~cm$^{-2}$~s$^{-1}$
(fixing N$_H$ at the Galactic value, 3.89$\times 10^{20}$atoms~
cm$^{-2}$). The photon index of the inner jet knot is 
affected by the core background radiation, since the photon index of
the core emission is much harder with $\Gamma=1.19\pm 0.02$ and is
absorbed (Bechtold et al 2001).

\subsection{Comparison of X-ray and Radio Jets}

Figure 7a shows the {\it Chandra} X-ray image (color scale) smoothed
to $\sim 3\arcsec$ resolution in order to match the 1.4~GHz beam size
overlayed with the contours from the 1.4 GHz VLA radio map. The X-ray
and radio jet emission are co-extensive.  There is also a general
correspondence in the sense of alignment of the jet features, but it
is obvious that we do not have detailed agreement. The peak brightness
of the X-ray emission precedes the peak of the 1.4~GHz emission for
each knot by about $\sim0.8-2\arcsec$ (see Table 2) corresponding to
$\sim$8-20$h_{50}^{-1}$~kpc at the quasar redshift.

These 2.6$\sigma$ offsets are not due to astrometric differences
between the X-ray and radio images. The peaks of the X-ray and radio
core emissions have been aligned, and the aspect uncertainty in the
X-ray image is in any case of order $\sim 0.5\arcsec$.

The offsets can be critical in constraining jet emission models (see
Sec. 4).  However, the analysis of the emission processes are
complicated, because there is no detailed correspondence between the
radio and X-ray emitting regions. Table 1 contains the total flux
densities for each knot measured in radio and X-rays independently. In
Table 3 we included only the flux densities which were obtained for
the same regions in radio and X-rays.  We first defined the X-ray
regions and then measured the corresponding (to that region) radio flux
density. We used these flux densities in calculating the model
parameters.

Figure 7b presents the smoothed X-ray and 8.4~GHz data at
$\sim$0.8$\arcsec$ resolution. This higher resolution radio emission
shows more complex structure than the X-ray emission at this
resolution.  There is no radio emission at this frequency associated
with the continuous X-ray jet emission coming directly from the core
nor is knot A detected at this frequency, while it is the strongest
X-ray knot. The two outer knots, B and C, dominate the radio emission.
The X-ray emission in knot C is more compact and narrow along the jet
direction than the radio.  Also for knot B and C the strongest X-ray
emission is upstream of the bulk of the radio emission. 


Overall the X-ray emission declines with the distance from the core
and is strongest for knot A, which is weak and barely detected in
radio (see Figure 4). 

\begin{figure}
\epsscale{1.0}
\plottwo{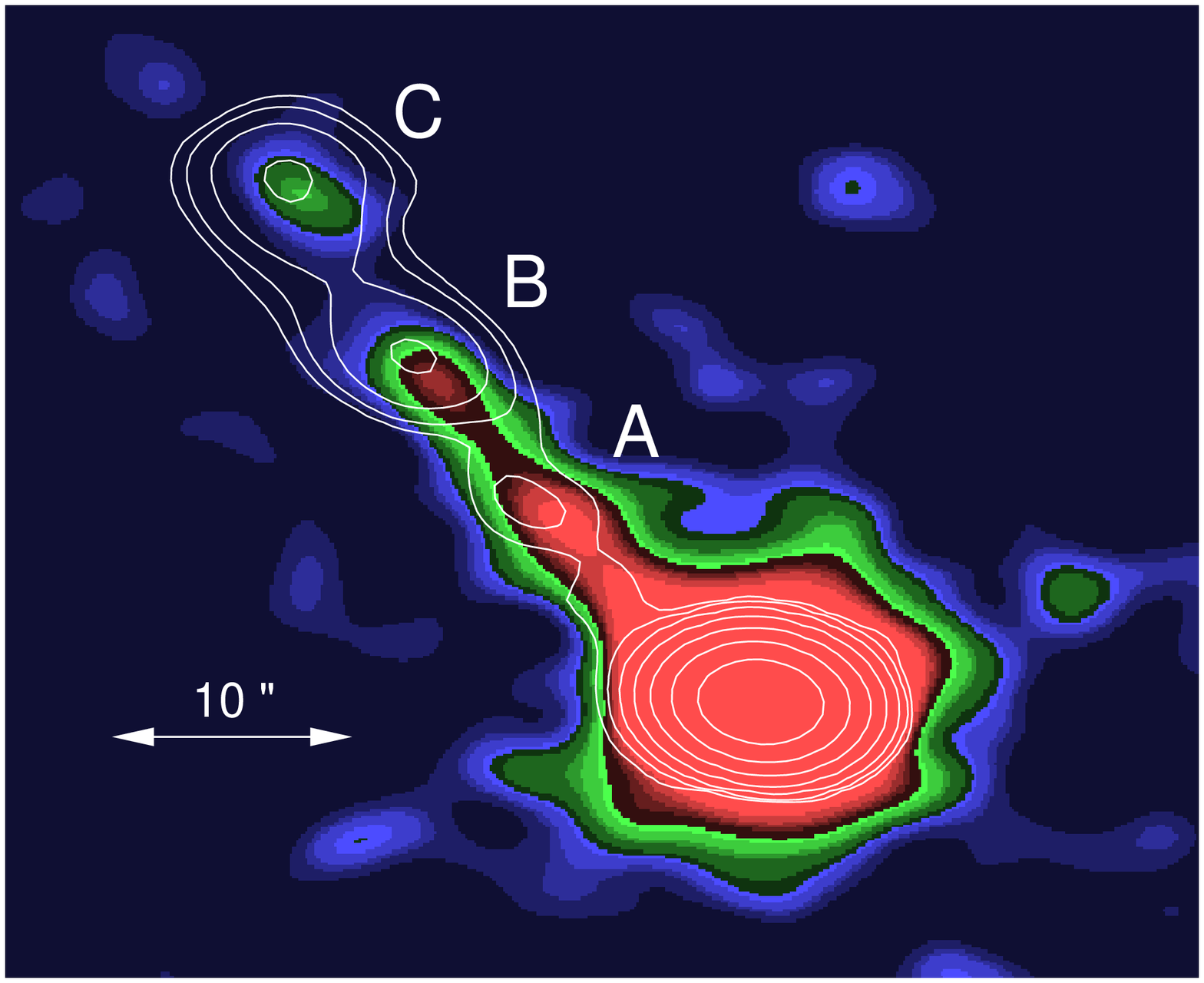}{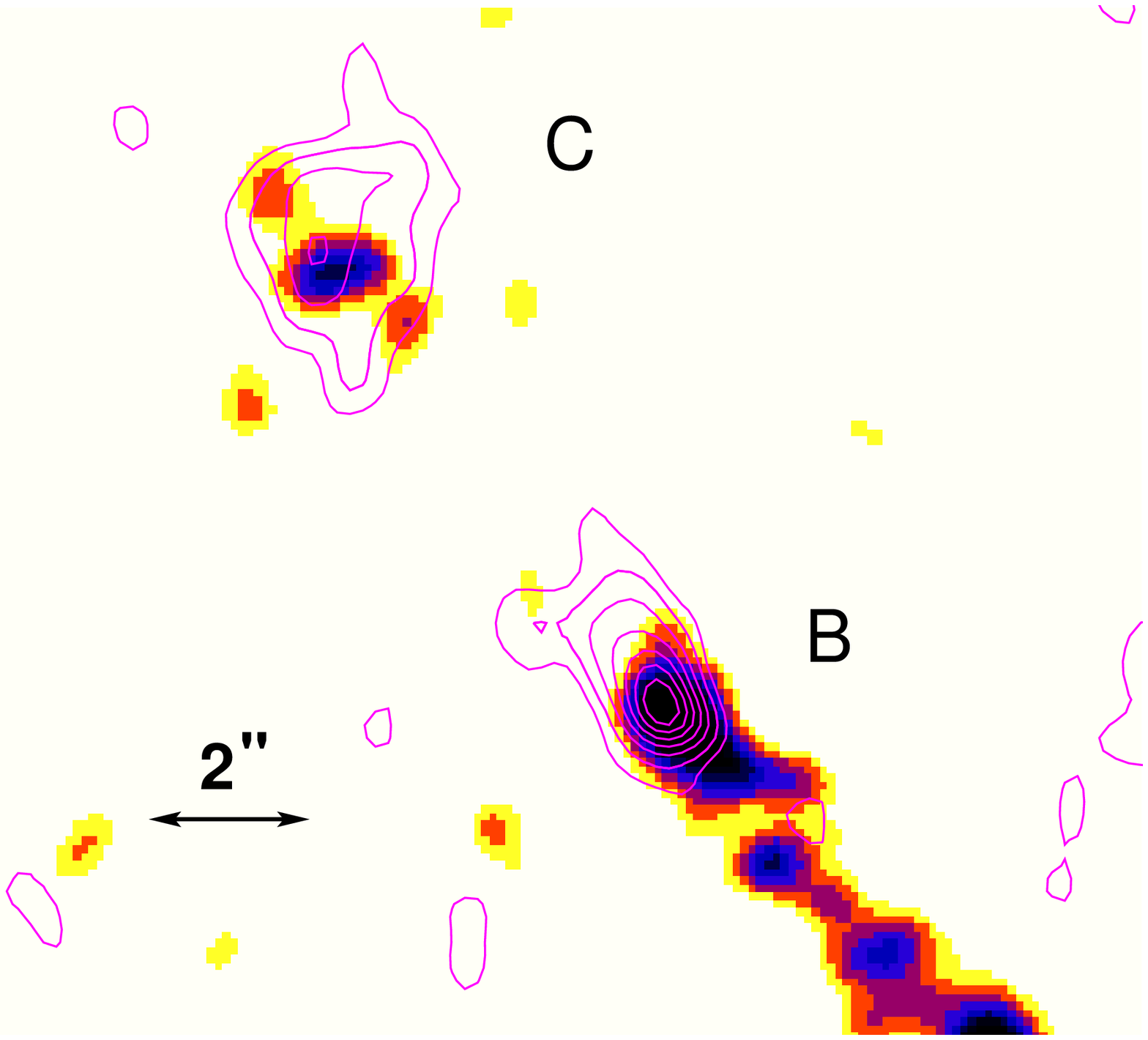}
\caption
{\small a) X-ray color map overlayed with the low resolution 
1.4GHz radio contours. X-ray data have been smoothed to match the
radio beam size. The X-ray peak intensities precede the radio for each
knot.  b) An overlay of the Chandra image (color, smoothed to
$\sim 1^{\prime\prime}$) and the 8.4 GHz contours.  The color scale
ranges from 0.0002 to 0.002 of the peak X-ray brightness at the quasar
nucleus of PKS1127-145 (which is just off the figure to the lower
right).  The contour levels are linear from 1 to 10~$\times$~0.2
mJy/beam and the restoring beamwidth is
0.89$^{\prime\prime}~\times~0.82^{\prime\prime}$ in PA=65$^{\circ}$.}
\end{figure}

\begin{figure}
\epsscale{0.85}
\plotone{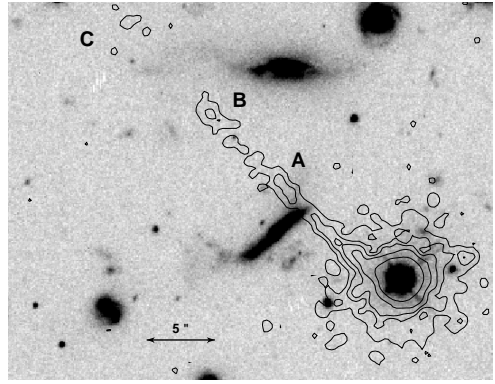}
\caption{\small Overlayed X-ray contours on the HST/WFPC2 image (F814W
filter).}
\end{figure}

\begin{figure}
\epsscale{0.9}
\plottwo{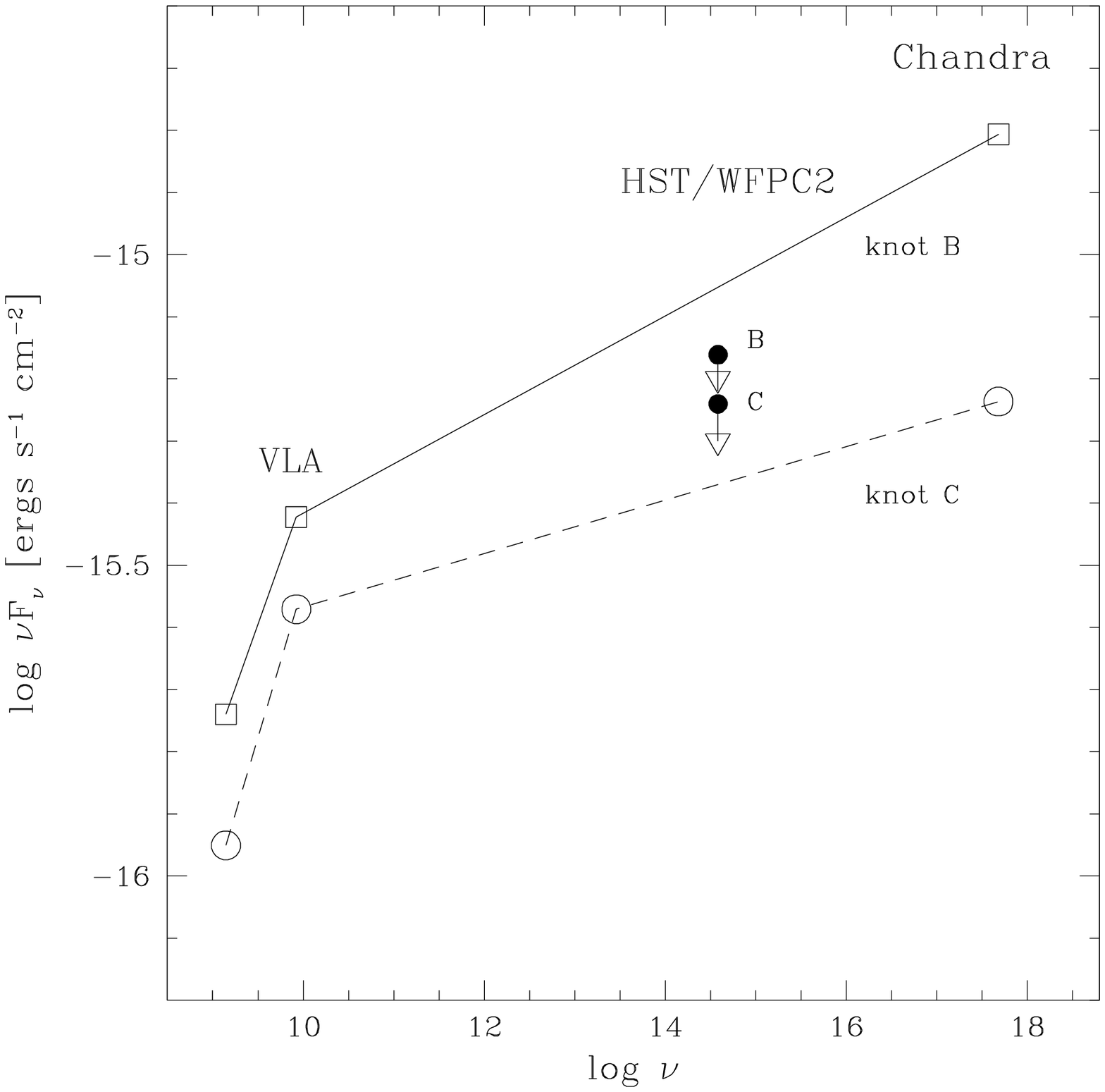}{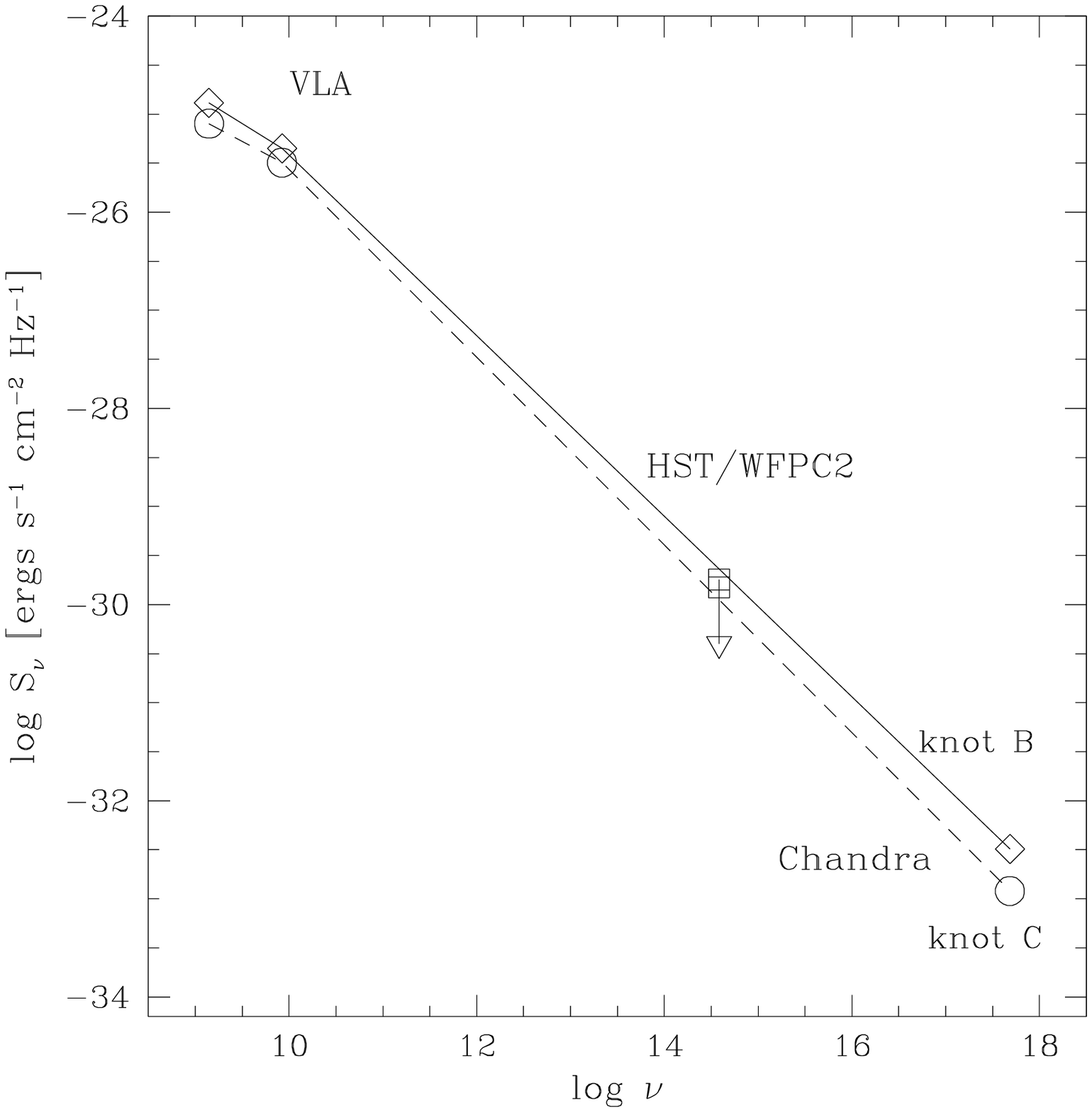}
\caption{\small 
a) Spectral energy distribution ($\nu F_{\nu}$) for knots B and C. The
HST/WFPC2 optical 3$\sigma$ upper limits for knot B and C are
indicated with the arrows. b) Flux density for knot B and C. The
optical upper limit is indicated with an arrow.}
\end{figure}

\subsection{Search for optical jet counterparts}

Figure~8 contains the HST/WFPC2 image with X-ray contours overlayed, showing
that there is no obvious optical counterpart to the X-ray/radio jet in
\pks.  This image also shows that the field surrounding
\pks\ is rich with galaxies.  Nestor et al. (2001) obtained Keck 
spectroscopy of the nearby galaxies and found a redshift of 0.312 for
several of them, indicating the presence of a foreground galaxy group
or cluster at this redshift, although there is no obvious optical
group or cluster present in our data.

As seen in Figure~8, detection of an optical counterpart to the X-ray
jet (and knot A in particular) is hampered by structure in the
foreground galaxies.  Nevertheless, we can calculate useful upper
limits to optical emission for knots B and C.  This was done by
summing the counts in precisely the same source regions used to
extract the radio and X-ray fluxes (see Table~3).  The regions for
background subtraction were boxes surrounding the knot regions, with
the source and any point sources excluded.  For both knots, the net
counts are consistent with zero, and we report 3-$\sigma$ upper
limits.  For knot B,
$S_{\nu}(3.83\times10^{14}Hz) < 1.8 \times 10^{-30}$\,erg\,cm$^{-2}$\,s$^{-1}$\,Hz$^{-1}$,
and for knot C,
$S_{\nu}(3.83 \times10^{14} Hz) < 1.5 \times 10^{-30}$\,erg\,cm$^{-2}$\,s$^{-1}$\,Hz$^{-1}$.
The limits in instrument counts were converted to fluxes using the formulas and
tables Chapter 6 of the WFPC2 Instrument Handbook Version 6.0 (Biretta et
al. 2001)

\section{Jet X-ray Emission Processes}

Several emission processes are potentially responsible for the jet
X-ray emission: synchrotron, synchrotron self-compton (SSC), external
inverse compton (EIC) models, and thermal.  In this section we attempt
to reconcile the observed data with each of these model processes and
discuss the results. 

Table 3 summarizes the input parameters used in the modeling of the
jet emission.  Because the X-ray and radio emission regions of \pks\
are not quite the same we chose the regions based on the X-ray
emission first and then measured the radio flux densities for those
regions. Table 3 defines the regions for each knot. We also included a
10.2$\arcsec$ box region to measure the continuous emission along the
jet in the vicinity of knot~A.

\subsection{Thermal radiation}

We can rule out the possibility of the jet X-ray emission arising from
hot thermal plasma associated with the knots, since the required
particle densities ($>$0.1-0.3 atoms cm$^{-3}$) and total mass
(2-20$\times 10^{10}$M$_{\odot}$) required to generate the observed
X-ray luminosities are very unlikely to be found at great distances
from the quasar core (see Table 3).

\subsection{Synchrotron Emission}

Synchrotron X-ray emission requires moderate magnetic fields (B$\sim
10^{-3}-10^{-5}$G) and highly relativistic particles ($\gamma \sim
10^7-10^8$, where $\gamma$ is the Lorentz factor of the relativistic
electrons) to be present at the emission site.  Such highly
relativistic electrons have short timescales for energy losses due to
synchrotron or Compton radiation ($\sim 10-10^3$ years depending on
the magnetic field) implying that the electron acceleration and
synchrotron emission must occur in the same region. Therefore the
radio and X-ray emission regions should be coincident to $\leq 10^3$
l-yr, if synchrotron or SSC emission is responsible for the
X-rays. However, the radio and X-ray peak brightnesses in \pks\, are
shifted by large distances ($\sim 10-20$~kpc) and the regions of the
strong X-ray emission do not exactly coincide with the radio emitting
regions. There are also regions of X-ray emission with no
corresponding radio emission (see Figure 7). This argues against a
simple synchrotron model and would require that the electron spectrum
is flat with the index $p<2$ (where n$_{\gamma} = n_0
\gamma^{-p}$d$\gamma$, $p=2\alpha+1$, and $S_{\nu}\sim {\nu}^{-\alpha}$).

Synchrotron emission results in a continuous spectrum between radio
and X-ray bands, therefore a second critical constraint for this model
comes from the deep optical imaging. Our HST/WFPC data give upper
limits for the emission from knots B and C (see Figure 9). This limit
is in agreement with a simple extrapolation between the radio and
X-ray emission consistent with the synchrotron emission for knot C,
but requires a clear spectral break in the optical band for knot B
(Fig. 9). A presence of this break suggests that a simple synchrotron
model cannot explain the X-ray emission. A flatter spectral component
is required if X-rays were to be a result of synchrotron emission for
knot B. Also more complex synchrotron models are possible.

\subsection{SSC emission}

In the SSC process the synchrotron photons generated by the
relativistic jet electrons are Compton scattered by the same
electrons, gaining a factor of $\gamma^2$ in energy. If the mean
electron energy is of order $\gamma \sim 10^4$ then photon energies
(frequency) are increased by a factor of 10$^8$, so radio photons
(10$^9$-10$^{11}$Hz) become X-ray photons (10$^{17}-10^{19}$Hz).  The
observed energy density of the synchrotron photons in large scale jets
is usually much too small to produce the observed X-rays for magnetic
field strengths calculated for equipartition except in a few terminal
hot spots (e.g. Cygnus A, Harris et al. 1994, Wilson et al 2001;
3C123, Hardcastle, Birkinshaw \& Worrall 2001).  The SSC model for
X-ray jet emission in \pks\ can be ruled out because it predicts much
lower (by factors of hundreds or more) X-ray luminosities than we
observe (see $S_{(obs)}/S_{(SCC)}$ in Table 3).

\subsection{EIC/CMB model}

The remaining possibility is external inverse Compton emission.  In
EIC models, the external photons are Compton scattered off the
relativistic particles in the jet.

The source of external photons could be provided by the AGN itself,
scattered light of the host galaxy or the Cosmic Microwave Background
(CMB).  If the jet extends over hundreds of kiloparsec from the
nucleus, the external photons related to the AGN become less important
than the background photons. Similarly, far from the optical galaxy
(150-300$h_{50}^{-1}$(sin$\theta)^{-1}$~kpc) ambient starlight will
have lower energy density than the Cosmic Microwave Background (CMB)
radiation

We therefore considered CMB photons scattered off the radio emitting
relativistic electrons for the source of the X-ray emission, as
suggested by Cellotti et al (2001) and Tavecchio et al (2001) for the
PKS~0637-752 jet (Schwartz et al 2000). If the jet was moving with
significant bulk velocity, then in the jet frame the electrons
experience an enhanced photon energy density (by a factor $\Gamma^2$,
where $\Gamma$ is the bulk Lorentz factor of the jet). IC losses then
dominate over synchrotron losses. The photon energy density of the CMB
at the redshift of \pks\ is $\sim$9.75$\times 10^{-12}$erg cm$^{-3}$
and the observed radio and X-ray flux ratios do not require high bulk
velocities. Relatively low Lorentz factors, $\Gamma \sim 2-3$ are then
needed to obtain the observed X-ray flux (see Table 3, the model
parameters were calculated using the formalism of Harris \&
Krawczynski 2002).

Note that the jet beaming model does not require the presence of {\it
highly} relativistic synchrotron electrons, since $\gamma < 10^3$
electrons can scatter the CMB photons to the X-ray band. As a result
the X-ray jet emission can be generated from an aged electron
population resulting in much less radio emission than in the case of a
stationary source.  For \pks\ electrons which scatter CMB photons to
$\sim 1$keV produce $\sim$1~MHz radio emission ($\Gamma = 1-3$). This
model allows then for the presence of the observed X-ray emission
without detectable radio emission.

Both outer knots (B and C) have 8.4~GHz emission, while the jet
continuous radio structure is not very prominent. 
The EIC/CMB X-ray jet emission indicates a presence of low energy
electrons along the jet, but their energy is too low for synchrotron
emission at the observed frequency.  The 8.4 GHz structure suggests a
production of higher energy electrons at the emission sites, thus a
presence of a shock.

The observed radio emission could well be lobe emission rather than
jet emission (see Sec.3.2).  Although we used only radio flux from the
observed areas emitting X-rays in our beaming calculations, if all of
the radio emission associated with the outer features (A, B, and C)
arises from a lobe or cocoon around the (X-ray) jet, the actual
beaming parameters reported in Table 3 would not be accurate since we
would have no information on the synchrotron component of the jet
emission.  With our failure to detect any counter jet (or counter
lobe), we would have a source with much the same problem as that
encountered in explaining the larger radio structures of the 3C~273
jet: if the predominant radio structure is a cocoon around the
optical/X-ray jet, why is the lobe on the other side not detected?
can one sided lobe be due to the voids or filaments in the IGM?

\section{Discussion.}

\subsection{Jet models.}

We can consider the morphology of X-ray jets in terms of brightness
contrast along the jet.  At one extreme is the inner X-ray jet of
Pictor A which appears essentially continuous and smooth (Wilson,
Young \& Shopbell 2001, Marshall et al 2001); and at the other extreme
is the jet of M87 for which knot A dominates to such an extent that
its emission rivals that of the core (Biretta, Stern \& Harris 1991).
While it is true that the {\it Chandra} physical resolution of the
\pks\ jet is only 8-10~kpc (cf. $\sim$1~kpc for Pictor A, Wilson et al
2001, and $\sim$100~pc for M87 jet, Marshall et al 2001), the X-ray
jet emission of \pks\, has only modest brightness enhancements (factor
$\sim 2$, Fig.3), perhaps not really sufficient to define discrete
knots.  Some of the X-ray knot structure may be due to patchy
foreground absorption. The foreground galaxy (Fig.8) may absorb X-rays
sufficiently to artificially separate knot A from the inner jet.  A
column density of 10$^{22}$~atoms cm$^{-2}$ would depress the observed
flux by the necessary factor of 2 (for an assumed photon index of
$\Gamma=1.5$), although this column seems too large for such an
extended region, where column densities of $\le 10^{21}$~atoms
cm$^{-2}$ are more likely (e.g. in damped Lyman alpha systems, Wolfe
1988). For a steeper photon index ($\Gamma>2.5$) a column density of
$10^{21}$~atoms~cm$^{-2}$ is enough to obtain the required reduction
in flux. Note that \pks\, has damped Lyman alpha system in the optical
spectrum (Lane et al 1998).

For knots A and B, the brightest parts of the radio emission occur
downstream of the peak X-ray brightness and the brightest radio region
of knot C appears to form an outer cocoon around the X-ray knot C
(Figure 6b).  In both outer knots (B and C) the radio emission extends
further out along the jet in a manner reminiscent of the end of the 3C
273 jet (Marshall et al 2001). Knot B lies at the end of a continuous
region of X-ray emission.

Knot A is not detected at 8.4~GHz. The steep implied radio spectrum
suggests that it has a different electron energy distribution than the
other two knots, B and C. This could be a result of a weaker shock
acceleration than in the other two knots. Also it is possible that
this is a location of past shock activity and we now see only low
energy particles with the longer lifetime than the higher energy
particles. The energy losses are mostly due to the synchrotron and
Inverse Compton process and for electrons with $\gamma < 10^4$ these
timescales are longer than 10$^4-10^5$ years.

We concluded in Sec.4 that thermal and SSC models are unlikely. While
a synchrotron model is possible for knot C, a simple synchrotron
emission cannot explain knot B emission.  Instead EIC beaming models
using the CMB as a source of photons can easily accommodate the
observations.  The modest beaming parameters are a result of the fact
that even for stationary sources, the energy density of the microwave
background increases with redshift as $(1+z)^4$ and so is 22.8 times
greater at the \pks\ redshift than locally.  If EIC is responsible for
the X-ray emission, then the average ratio of X-ray to radio
luminosities for jets should increase with redshift. Such statistical
studies will be possible in the future when more X-ray jets become
available from {\it Chandra} observations.

\pks\, jet is the longest X-ray jet detected at high redshift. It
is difficult to compare its structure and emission properties to the
nearby jets discovered recently with Chandra.  Note that typical
physical sizes of the resolved X-ray structures in our source are of
the order of 10-15~kpc (projected on the sky), much larger than the
structures seen in 3C~273 or M87 (Marshall et al 2001a,
2001b). Synchrotron emission is most likely responsible for the jet
X-ray emission in these two nearby sources, but their projected X-ray
jets are rather small in comparison to \pks\, (0.1-70~kpc
vs. 330~kpc). Knot A in \pks\, is at $\sim 100$~kpc from the nucleus,
while the entire 3C~273 jet is well within that distance.  It is not
surprising that emission processes might be different.

At high redshift the energy density of the CMB increases and enhances
IC losses relative to synchrotron losses.  This allows for X-ray
emission even if there is no steady supply of highly relativistic
synchrotron electrons. Thus at high redshift X-ray emission at large
distance from the nucleus can be easily dominated by EIC, while at
lower redshift synchrotron or SSC dominate. At intermediate redshift
of PKS~0637-752 (z=0.653) the dominant process is not clear, although
equipartition argument suggests that EIC/CMB wins (Tavecchio et al
2000, Celotti et al 2001). \pks\, is probably the most promising
example of the X-ray emission arising from the interaction between CMB
photons and the jet plasma.

\subsection{GPS Sources.}

The detection of a large scale ($>$300~kpc) X-ray jet in \pks\ has
some interesting implications for our understanding of the nature of
Giga-Hertz Peaked Spectrum (GPS) sources.  These powerful radio
emitters ($\log P_{1.4} > 25$ W~Hz$^{-1}$) are characterized by convex
radio spectra which peak around $\sim 1- 5$~GHz.  They have a similar
morphology to the large (100~kpc -- 1~Mpc) FRI/FRII radio galaxies but
are much more compact with projected total sizes of less than $<
1$~kpc).  Some GPS sources exhibit very faint extended radio emission
on scales larger than the size of the host galaxy (Baum et al 1990,
Stanghellini et al 1990, O'Dea 1998, Fanti et al 2001), but only a few
GPS galaxies show faint radio structures on Mpc scales (Schoenmakers
et al 1999).

There are three different explanations of the nature of GPS sources.
In one scenario the jet cannot escape the dense environment and the
radio source is confined to a small central region (``frustrated''
radio sources e.g. van Breugel, Miley \& Heckman et al 1984, O'Dea,
Baum \& Stanghellini, 1991).  This is supported by X-ray observations
which show associated absorption in several GPS quasars (Elvis et al
1994).  Also Marr, Taylor and Crawford (2001) argue that the turn-over
in the spectrum in 0108+388 can be due to free-free absorption if the
gas is non-uniform within the central tens of parsecs.

In the second scenario the compact morphology of GPS sources is a
result of their small age: GPS sources may be young versions of the
large scale radio galaxies, so we observe them at an earlier stage of
their expansion (Phillips \& Mutel 1982). Supportive evidence for the
young age of the GPS sources comes from the propagation velocities of
the hot spots measured for several GPS sources (Owsianik \& Conway
1998, Owsianik, Conway \& Polatidis 1999, Tschager et al 2000). The
average expansion velocities are typically of order 0.2$h^{-1}c$
indicating kinematic age of $\sim 10^3$ years.  In this model the
large scale extended emission is usually interpreted as a relic of
much older previous activity (0108+388 Baum et al 1990, Owsianik,
Conway \& Polatidis 1998, Marr et al 2001; Stanghellini et al 1990,
Schoenmakers et al 1999), while the GPS ``core'' is young, maybe
rejuvenated by the inflow of new matter from a recent merger or
accretion disk instability (Siemiginowska, Czerny \& Kostyunin 1996,
Mineshige \& Shields 1990).

The simple evolution scenario in which GPS sources evolve with
constant luminosity and advance speed into large radio sources fails
because of their large relative number (O'Dea 1998). They form at
least 10$\%$ of the bright radio source population, while based on
their lifetime they should be much less numerous (only 0.1$\%$).  To
explain the large number of GPS sources Reynolds \& Begelman (1997)
(also O'Dea \& Baum 1997) suggested that extragalactic radio sources
undergo intermittent periods of activity (with a duty cycle of 10$^7$
years): the radio jets and core are bright when the AGN is ``on'', but
when it turns off, the jet radio emission fades rapidly, although the
jet structure remains intact and the matter keeps expanding.  The
large scale (100~kpc--1~Mpc) radio emission is a factor of $10^2-10^3$
weaker than at the time when the source was ``on'', and so is hard to
detect.  Instead, the large scale X-ray emission reported here
highlights the fact that if there are enough relativistic particles in
the expanding plasma then the interaction between the particles and
CMB photons results in the detectable X-ray emission. The X-rays can
then be used for studying large scale ``fossil'' jet structure at high
redshifts.

If GPS radio sources are young, with a short repetition timescale, we
can use them to test models of quasar evolution. The most common
scenario is one in which the quasar activity is linked to a merger
event (e.g. Mihos \& Hernquist 1996). This scenario could be tested by
studying the rates of merging galaxies required by the number of newly
rejuvenated GPS sources. Alternative physical models which link quasar
activity with instabilities in the accretion disk (Hatziminaoglou,
Siemiginowska \& Elvis 2001, Siemiginowska \& Elvis 1997) give a
repetition timescale which scales roughly with the black hole mass (a
timescale of 10$^6$ years corresponds to $\sim$10$^8$M$_{\odot}$ for
L$_{Edd} \sim 10^{46}$ erg~s$^{-1}$).  A third option (Ciotti \&
Ostriker 2001) is a feedback mechanism in which the accreting source
heats the ambient gas to the point at which the accretion stops (when
the Compton temperature of the emitted radiation exceeds the virial
temperature of the galactic gas). Accretion restarts after the gas has
cooled. The outburst phase is short in comparison to the cooling
phase. Studying the extended structures of the GPS sources can provide
constraints on the length of the outburst phase and potential test for
this model.

\subsection{\pks\, as a GPS source.}

GPS classification is based on the radio spectral properties, which
results in a heterogeneous sample.  GPS quasars often have a core-jet
VLBI (milliarcsec) morphology, different from GPS radio galaxies,
which have double components and a core (Stanghellini et al 2001). The
core-jet morphology in the GPS quasars may suggest that the convex
radio spectrum is due to beamed jet emission. Thus, often GPS quasars
can be identified with blazars or core-dominated quasars, in contrast
to GPS radio galaxies.

\pks\ has the properties of a GPS quasar (Stanghellini et al 1996), 
with low optical polarization (Impey \& Tapia 1990), and it is not an
optically violently variable source (Pica et al. 1988; Bozyan,
Hemenway, \& Argue 1990).  VLBI observations show the presence of a
double component with the parsec scale jet on one side (Wehrle et al
1992, Stanghellini et al. 1998). The core emission also dominates in
X-rays and the X-ray knot emission is only of the order of 0.1-0.2\%
of the core emission.  Although it has been tentatively identified as
an EGRET source, suggesting a blazar nature, the radio source lies
outside the 99$\%$ EGRET error region (Hartmann et al 1999).  The
optical spectrum does not provide any signatures of a blazar nature,
rather it has a typical quasar continuum shape with broad emission
lines and absorption features (Bechtold et al 2001).

The classification of \pks\, as a GPS source, although the source has
been included in the O'Dea (1998) sample, is still not certain.  The
age of the GPS source can only be determined through propagation
velocities of the hot spots within the core, as has been done for
several nearby sources (Owsianik et al 1999, Fanti
2000). Unfortunately this technique is impractical for high redshift
sources.  VLBI measurements could monitor variability of the central
components and characteristic timescales within the central region.
If the source is a blazar, then the jet indicates beaming of the
central core. Improved high energy X-ray (E$>50$~keV) and $\gamma$-ray
observations could provide clues to a possible blazar nature of the
source.

\section{Summary}

Our results can be summarized as follows:

\begin{itemize}
\item We have observed \pks\, for 27 ksec with {\it Chandra} and
detect an X-ray jet with a minimum length of 330$h_{50}^{-1}$~kpc.

\item Comparison of X-ray, optical and radio data of the \pks\, jet 
rules out thermal emission, SSC and a simple direct synchrotron
emission as primary source of the X-ray jet emission.

\item Inverse Compton scattering off CMB photons with moderate jet
bulk velocities can readily accommodate the observations and so is the
most probable emission mechanism.

\item The EIC/CMB process is especially effective at high redshift
because of the $(1+z)^4$ scaling of the CMB. X-rays from EIC can trace
low energy ($\gamma \sim 10^3$) population of particles which are not
detectable in the radio band and also delineate the ``fossil jet''
structure.

\end{itemize}

\acknowledgments

P. Kronberg and M. Cummins kindly faxed us the relevant pages of
Rusk's thesis and C. Carilli provided archival VLA data.  We also
thank C. Carilli and F. Owen for helpful advice on VLA data
calibration. AS thanks A. Celotti, M.Sikora and C.Stanghellini for
discussion on jet models and GPS sources.  We are grateful to Holly
Jessop for helping in creating the final Chandra images. We also thank
an anonymous referee for valuable comments.  This research is funded
in part by NASA contracts NAS8-39073.  Partial support for this work
was provided by the National Aeronautics and Space Administration
through Chandra Award Number GO-01164X issued by the Chandra X-Ray
Observatory Center, which is operated by the Smithsonian Astrophysical
Observatory for and on behalf of NASA under contract NAS8-39073.
Support for proposal HST-GO-09173 was provided by NASA through a grant
from the Space Telescope Science Institute, which is operated by the
Association of Universities for Research in Astronomy, Inc., under
NASA contract NAS 5-26555.



\begin{table*}[ht]
\caption{Jet and Individual Knots Measurements}
\smallskip
\begin{scriptsize}
\label{tab}
\begin{tabular}{lcccccccccccc}
\hline
\hline\noalign{\smallskip}
&&&X-rays &&&&RADIO \\
\hline\noalign{\smallskip}
 & Total & Net {\small (0.3-7)} & {\small S(0.1-2)$^a$}&
{\small H(2-10)$^a$} & HR$^b$  &F$^{c,d}$ &$S^{e}$ & $S^{e}$ &
$\alpha_r ^h$ \\
 &  Cts&Cts&Total(Net)& Total(Net) &&2-10 keV& 1.4 GHz & 8.4 GHz \\
\hline
\hline\noalign{\smallskip}
Core$^f$ &  16,573 &16,339.6${\pm 2.1}$ & 10,722(10,719.1) & 5,777(5,774.3) & -0.30 &241.3&5463 & 3427 &0.26\\ 
Jet total & 368 & 273.7${\pm 22.1}$& 250(231.1) & 71(55.1)& -0.61& 5.19${\pm 0.33}$& & \\ 
Inner knot & 546 & 56${\pm 21}$& 348(43.7) & 195(11.3) &
-0.59&1.27${\pm 0.43}$ & &4.6\\ 
A&50 & 41.8${\pm 7.2}$& 37(35.8)& 11(10.2) & -0.56 &0.93${\pm 0.14}$ &6${\pm 1}$ & $<1$& 1.2\\ 
B &46 & 34.4${\pm 5.9}$& 29(27.5)& 9(7.7)& -0.56&0.75${\pm 0.12}$&44${\pm 2}$ & 7.7${\pm 0.5}$& 0.82\\ 
C &34 & 21.0${\pm 4.7}$& 17(15.3)& 9(7.1)& -0.37 &0.51${\pm 0.12}$
&52${\pm 2}$ &10.7${\pm 0.5}$& 0.86 \\
\hline
\hline\noalign{\smallskip}
\end{tabular}

{\small 

$^a$ Net counts are shown in the brackets.

$^b$ Hardness ratio HR=H-S/H+S; 1$\sigma$ errors equal to
$\sigma$=0.17,0.20,0.25 for knots A,B,C and $\sigma$=0.53 for the inner knot.

$^c$ Galactic absorbing column assumed in all models: $3.89 \times
10^{20} cm^{-2}$;

$^d$ 2-10 keV flux in units of $10^{-14}$ erg s$^{-1}$ cm$^{-2}$,
assuming a photon index  $\Gamma$= 1.5 for all components.

$^e$ Radio flux density in mJy
($10^{-26}$ergs~cm$^{-2}$s$^{-1}$Hz$^{-1}$) with estimated errors;
the error for the core arises solely from the absolute flux density
scale which should be good to 2\%.

$^g$ The size of the region box used to measure the radio flux density
for each knot
at 1.4 GHz:  A: 6$\arcsec$x6$\arcsec$, B:9$\arcsec$x7$\arcsec$
C:10.$\arcsec$5x7.5$\arcsec$; and at 
8.4 GHz: A: 5$\arcsec$x5$\arcsec$, B:4.0$\arcsec$x4.6$\arcsec$ C:
5$\arcsec$x6$\arcsec$, and inner knot: 1.9$\arcsec$x1.3$\arcsec$.

$^h$ radio spectral index; S$_{\nu}\sim \nu ^{-\alpha_r}$; 1$\sigma$
errors equal to 0.5, 0.15, 0.12 for knots A,B,C respectively. Spectral
index measurements were made using the new set of data obtained at
4.86~GHz. We used the similar UV coverage in the measurements of the
spectral index. The detailed analysis of the new radio data will be
reported elsewhere.}
\end{scriptsize} 
\end{table*}


\begin{table*}
\caption{\small Distance$^a$ of Knots From Core}
\smallskip
\begin{tabular}{lcccccc}
\hline
\hline\noalign{\smallskip}
Knot &r(X)$\arcsec$ & r(1.4)$\arcsec$ & $\Delta_{(R-X)}\arcsec$\\
\hline\noalign{\smallskip}
A  & 10.2 & 12.3& 2.1 \\
B  & 17.9 & 19.3& 1.4 \\
C  & 27.0 & 27.8& 0.8 \\
\hline\noalign{\smallskip}
\hline
\hline\noalign{\smallskip}
\end{tabular}

{\small
\parbox{2.5in}{
$^a$ Distance measured from core to peak brightness in knot.  X-ray
distance uncertainty is 0.3$\arcsec$ and 1.4~GHz uncertainty is
1$\arcsec$ }}
\end{table*}


\begin{table*}
\begin{scriptsize}
\caption{Model Parameters.}
\smallskip
\begin{tabular}{lcccccccccc}
\hline
\hline\noalign{\smallskip}
ITEM &		INNERJET &	10.2"BOX$^a$&	A	&	B	&	C	&	units\\
\hline
\hline\noalign{\smallskip}
	INPUT &&&&&& \\
\hline\noalign{\smallskip}
Region Size &	0.65x1.3&	1.2x10.2&	0.7x2.6&	0.9x2.8&	0.9x2.4&	arcsec\\
Distance$^c$  & 2. & 0.6 & 9.5 & 17.3& 25.9& arcsec \\
log Volume&	66.4	&	69.3	&	68.2&		68.5	&	68.425&		cm$^3$\\
$\alpha_r^b$	&	1&		1.33&		0.76&		0.92&		0.94&\\
S(8.4~GHz)	&	4.6&		&		$<$0.5&		4.5&		3.2&		mJy\\
S(1.4~GHz)	&	&		6&		2&		13&		8&		mJy\\
F$_x$(2-10~keV)	&	5.26&	14.6&	5.0&
4.0&		2.0&		10$^{-15}$ ergs cm$^{-2}$s$^{-1}$\\
S$_x$(2~keV)	&	0.65&	&	9.5& 7.8&		2.9&		10$^{-16}$ ergs cm$^{-2}$s$^{-1}$keV$^{-1}$\\
\hline\noalign{\smallskip}
   THERMAL BREMS$^d$&&&&&&& \\
\hline\noalign{\smallskip}
log Lx	&	43.9&		44.3	&	43.9&
43.8&		43.5	&	erg s$^{-1}$\\
n(e)&		2.12&		0.12&		0.23&		0.15&		0.12&		cm$^{-3}$\\
M&		0.38&		19&		3.4&		4.0
&	2.6&	 	10$^{10}$M$_{\odot}$\\
P&		26	&	1.4&		2.8&
1.9&		1.4&		10$^{-9}$dyne cm$^{-2}$\\
\hline\noalign{\smallskip}
   SYNCHROTRON$^f$&&&&&&\\
\hline\noalign{\smallskip}
B(eq)	&	 75&		5&		13.5&		16.4&		15.4&		$\mu$G\\
log Ls	&	44.1&		44.5	&	42.95&		43.8&		44.43&		erg/s\\
$\gamma_{18}$&	0.84 &		3&		2&	1.8& 1.8& 10$^8$\\
$t_{18}$&	 37&		310&		270&		237&		249&		years\\
\hline\noalign{\smallskip}
   SSC&&&&&& \\
\hline\noalign{\smallskip}
log Ls	&	43.6&		43.0	&	42.8 &43.5 &
43.4	&	erg s$^{-1}$ \\
B(eq)	&	 74&		8&		13.5&		16.3&		15.0&		$\mu$G\\
u($\nu$)	&	52&	0.12	&	0.48	&	0.91	&	1.5&	10$^{-13}$erg cm$^{-3}$\\
R=u($\nu$)/u(B)&	0.024&		0.004	&	0.007&		0.009&		0.016&\\
S(obs)/S(ssc)$^e$&	 708&		65,504&		14,400&		740&		313&\\
\hline\noalign{\smallskip}
   EIC$^j$ &&&&&&\\
\hline\noalign{\smallskip}
log L$x$ & & & 43.26 &	43.65&	43.49& erg~s$^{-1}$ \\
B(IC)$^g$&	3&		1.9&		1.9&		2.4&		2.8&		$\mu$G\\
$\Gamma$	&	 3.7&		1.7&		3.3&		2.4&		2.2&\\
R$^h$	&	 8.4&		31&		28&		11&		8&\\
$\theta^i$	&	15&		20&		 17&		24&		27&		degrees\\
\hline
\hline\noalign{\smallskip}
\end{tabular}
\end{scriptsize}

{\small

$^a$ the 10.2"box includes knot A and all regions are defined based on
the X-ray emission regions. 

$^b$ Synchrotron frequency range assumed for a given $\alpha_r$: $10^8-10^{11}$Hz.

$^c$ Distance of the beginning of the region from the core measured
along the jet.

$^d$ A typical X-ray temperature of the gas emitting in {\it
Chandra} band assumed to calculate the thermal gas
pressure: 4.4$\times 10^7$K

$^e$   S(obs)/S(ssc) is the 'failure factor' (observed/predicted
flux densities).

$^f$	Synchrotron frequency range covers the entire radio to X-ray
band: $10^8-10^{18}$Hz; $\gamma_{18}$ is the electron energy
responsible for 10$^{18}$~Hz emission; $t_{18}$ is the half-life time .
	
$^g$    B(IC) is the magnetic field required for IC/3k (no beaming).
	
$^h$	R is the ratio of IC to synchrotron losses in the jet frame.

$^i$	$\theta$ is the characteristic angle between the observer and
the jet. This is the maximum allowed angle in the model.}

$^j$  Synchrotron frequency range assumed in the EIC model:$10^6-10^{11}$Hz. 

\end{table*}

\clearpage

\end{document}